\documentclass[aps,prb,twocolumn,superscriptaddress,showpacs,floatfix]{revtex4}
\usepackage{graphicx}
\usepackage{amsmath}
\usepackage{amssymb}
\usepackage{color}
\usepackage[pdftex]{hyperref} 

\newcommand{\kk}{\mathbf{k}}
\newcommand{\pp}{\mathbf{p}}
\newcommand{\qq}{\mathbf{q}}

\newcommand{\PRL}{Phys. Rev. Lett. }
\newcommand{\PRB}{Phys. Rev. B }

\begin{document}
 
\title{Optical signatures of exciton-polarons from diagrammatic Monte Carlo}

\author{A. S. Mishchenko$^{1,2}$,  G. De Filippis$^3$, V. Cataudella$^3$, 
N.~Nagaosa$^{1,4}$, and H. Fehske$^5$}
\affiliation{
\mbox{RIKEN Center for Emergent Matter Science (CEMS), 2-1 Hirosawa, Wako, Saitama 
351-0198, Japan}\\
$^2$\mbox{NRC ``Kurchatov Institute'', 123182 - Moscow - Russia} \\
$^3$\mbox{Coherentia-INFM and Dip. di Scienze Fisiche - Universit\`{a} di Napoli Federico II - I-80126 Napoli, Italy}\\
$^4$\mbox{Department of Applied Physics, The University of Tokyo, 7-3-1 Hongo, Bunkyo-ku, Tokyo 113, Japan} \\
$^5$\mbox{Institut f{\"u}r Physik, Ernst-Moritz-Arndt-Universit{\"a}t Greifswald, 17489 Greifswald,
Germany} 
}

\begin{abstract}
We study the interplay of electron-electron and electron-phonon interactions in the course 
of electron-hole bound state formation for gapped solid state systems. 
Adapting the essentially approximation-free diagrammatic Monte Carlo method for the calculation 
of the optical response, we discuss the absorption of light in correlated electron-phonon systems 
for the whole interaction and phonon frequency regimes. 
The spectral function obtained by analytical continuation from the imaginary-time current-current 
correlation function demonstrates the dressing of excitons by a phonon cloud when the coupling 
the  lattice degrees of freedom becomes increasingly important, where notable differences show 
up between the adiabatic and anti-adiabatic cases. 
\end{abstract}
\maketitle


\section{Introduction.}

Optical properties of semiconductors and insulators are connected with 
interband transitions between the highest occupied  and lowest unoccupied bands, which can be 
probed by standard  light absorption measurements~\cite{ExpOA,Toyozawa_Book}.  
On the theoretical side, a full many-particle treatment of the resulting electron-hole systems 
is difficult, in particular if the (excitation) gap originates from strong electronic correlations 
or  coupling to the lattice degrees of freedom. The situation greatly simplifies  in the low-excitation regime, where an extremely small density of electrons and holes exist, and one can concentrate on the analysis of electron-hole pairing effects. Here the Coulomb interaction between conduction band electrons and valence band holes mainly triggers the formation of in-gap excitonic  bound states~\cite{Egri85,Ueta_Book}. 
Since the electron-hole pairs are electrically neutral and the coupling between carriers and optical phonons 
in polar compounds is electrical in nature, one naively would expect that the exciton-phonon interaction 
is weak. 
However, the coupling of carriers to phonons in non-polar solids is governed by deformation potential.
It was clear from the initial concept \cite{Bardeen_Shockley_1950} and then confirmed by
numerous studies, e.g. \cite{Def2,Def3}, that there is no fixed relation between the signs of the 
deformation potentials of valence and conduction bands. 
Thus, the exciton-phonon interaction can be significant.

One prominent example are robust exciton-polarons in (quasi-zero-dimensional) semiconductor quantum dots which strongly modify the photoluminescence (optical response) because the exciton and phonon states are entangled~\cite{Fomin98,VFB02}. Phonon-assisted electron-hole bound state formation  also quite often takes place in quasi-one-dimensional solids, such as polydiacethylene crystals~\cite{Wi83}. Particularly fascinating, the lattice seems to be involved in the phase transition to an excitonic insulator state which recently has been discussed for a number of novel materials at large exciton densities~\cite{HR68,PBF13,ZFBMB13,ZFB14}. For example, in semiconducting ${\rm Ta_2NiSe_5}$, the lattice structure changes from orthorhombic to monoclinic at the suggested excitonic instability~\cite{KTKO13}. A combination of excitonic and lattice instabilities has been made responsible for the observed (possibly chiral) charge-density wave  in the layered transition-metal dichalcogenide $1T$-${\rm TiSe_2}$~\cite{KMCC02}. Finally, in the intermediate-valent ${\rm TmSe_{0.45}Te_{0.55}}$ compound  the \mbox{(thermo-)} transport seems to be indicative of exciton-polarons as well~\cite{WBM04,WB13}.

First attempts to tackle theoretically the underlying tricky  exciton-polaron formation problem 
consider the exciton as a preformed structureless quasiparticle object~\cite{QDST,RP}. 
In addition, the frozen-phonon approximation was frequently used~\cite{Ueta_Book}, 
or simple variational approaches were exploited~\cite{Sumi}. 
The recently developed diagrammatic Monte Carlo  (DMC) technique~\cite{PS98,MPSS,exciton2001} 
seems to be especially suitable to address the long-standing exciton-polaron issue more 
seriously~\cite{expo2008}.

\section{Generic model}

Therefore, in the present paper, we generalize and apply the DMC to the calculation of 
light absorption by a coupled electron-hole-lattice system described by a generic  
gapped model system. 
In this way  we are able to provide exact numerical results for the optical response in the whole 
parameter range of Coulomb interaction, fermion-phonon coupling, and phonon frequency.

The two-band model Hamiltonian  reads
\begin{eqnarray}
H &=& \sum_\kk  \varepsilon_c(\kk) \, e^{\dagger}_\kk e_\kk + %
    \sum_\kk  \varepsilon_v(\kk)\, h_\kk \, h^{\dagger}_\kk + %
    \sum_\qq \omega_\qq \, b^{\dagger}_\qq b_\qq \nonumber \\
   &&- \sum_{\kk\qq} \left[ \frac{g_e(\qq)}{\sqrt{N}}
   e^\dagger_{\kk-\qq}e_{\kk} +
   \frac{g_h(\qq)}{\sqrt{N}}h^\dagger_{\kk-\qq}h_{\kk}
   \right] \left( b^\dagger_{\qq} + b_{-\qq} \right)\nonumber\\
   &&- \sum_{\pp\kk\kk'} \frac{U(\pp,\kk,\kk')}{N} \,
   e^{\dagger}_{\kk}h^{\dagger}_{\pp-\kk} h_{\pp-\kk'}
   e_{\kk'}\;,
\label{model}
\end{eqnarray}
where $e^\dagger_\kk$ [$h^\dagger_\kk$] creates an electron [a hole] in the conduction
[valence] band $\varepsilon_c(\kk)$ [$\varepsilon_v(\kk)$]. These electrons and holes  
feel an interband Coulomb attraction $U(\pp,\kk,\kk')$ that may cause the formation of
excitonic quasiparticles (electron-hole bound states) located in the gap between 
valence and conduction band. The coupling of the electrons [holes] to phonons created
 by $b^\dagger_\qq$ is parametrized by $g_e(\qq)$ [$g_h(\qq)$)], where $\omega_\qq$
is the energy of lattice vibrations in the harmonic approximation ($\hbar=1$). 
In Eq.~\eqref{model}, $N$ is the number of lattice sites; 
$\qq$, $\pp$, $\kk$, and $\kk'$ are momenta.

In a simplified model, preserving essential features of the phenomenon, 
we consider two tight-binding bands of a simple cubic lattice, 
\begin{equation}
\varepsilon_{c,v}(\kk) =\hat{E}_{c,v} \pm (W_{c,v}/6)
\sum_{\alpha=x,y,z} (1-\cos{k_\alpha})\,, 
\label{dispersion}
\end{equation}
having bandwidths $W_{c,v}$, where the energy of the 
valence-band top is set to zero ($\hat{E}_v=0$), i.e., the bottom of conduction band
gives the the direct band gap $E_g$ at $\kk=0$,  $\hat{E}_c=E_g$. Furthermore, we assume 
momentum independent Coulomb and electron-phonon interactions, $U(\pp,\kk,\kk')\equiv U$ and 
$g_{e,h}(\qq) \equiv g_{e,h}$, respectively, as well as dispersionless optical phonons 
$\omega_\qq \equiv \Omega$.
Then a dimensionless fermion-phonon coupling constants can be defined in the usual way as 
$$\lambda_{e,h}=2g_{e,h}^2 / (W_{e,h} \Omega) \;.$$
We set $W_{c,v}=3 E_g$, and use $E_g=1$ as energy unit.
The ground-state properties of the model~\eqref{model}--\eqref{dispersion} were previously 
analyzed by Burovski {\it et. al}~\cite{expo2008}.

\section{Absorption spectra at particular physical regimes}

Discussing the optical properties of exciton-polarons, we look at  
the  photon-absorption transition rate, which is  proportional to the spectral function
$A(\omega)$  that itself is related to  the real part of the optical conductivity $\sigma(\omega)$ 
by 
$$\mbox{Re}[\sigma(\omega)] = A(\omega) / \omega$$ 
\cite{Mahan_Book}.
To figure  out polaronic effects, we impose the normalization $\int_{0}^{\infty} d\omega  A(\omega)=1$.
In order to disentangle to some extent the complex interplay between Coulomb and 
electron/hole-phonon-coupling effects in the process of (exciton-polaron) bound-state formation, 
we analyze in what follows different characteristic situations for the three-dimensonal case.

\subsection{Pure Coulomb attraction}

\begin{figure}[b]
        \includegraphics[scale=0.32,clip=]{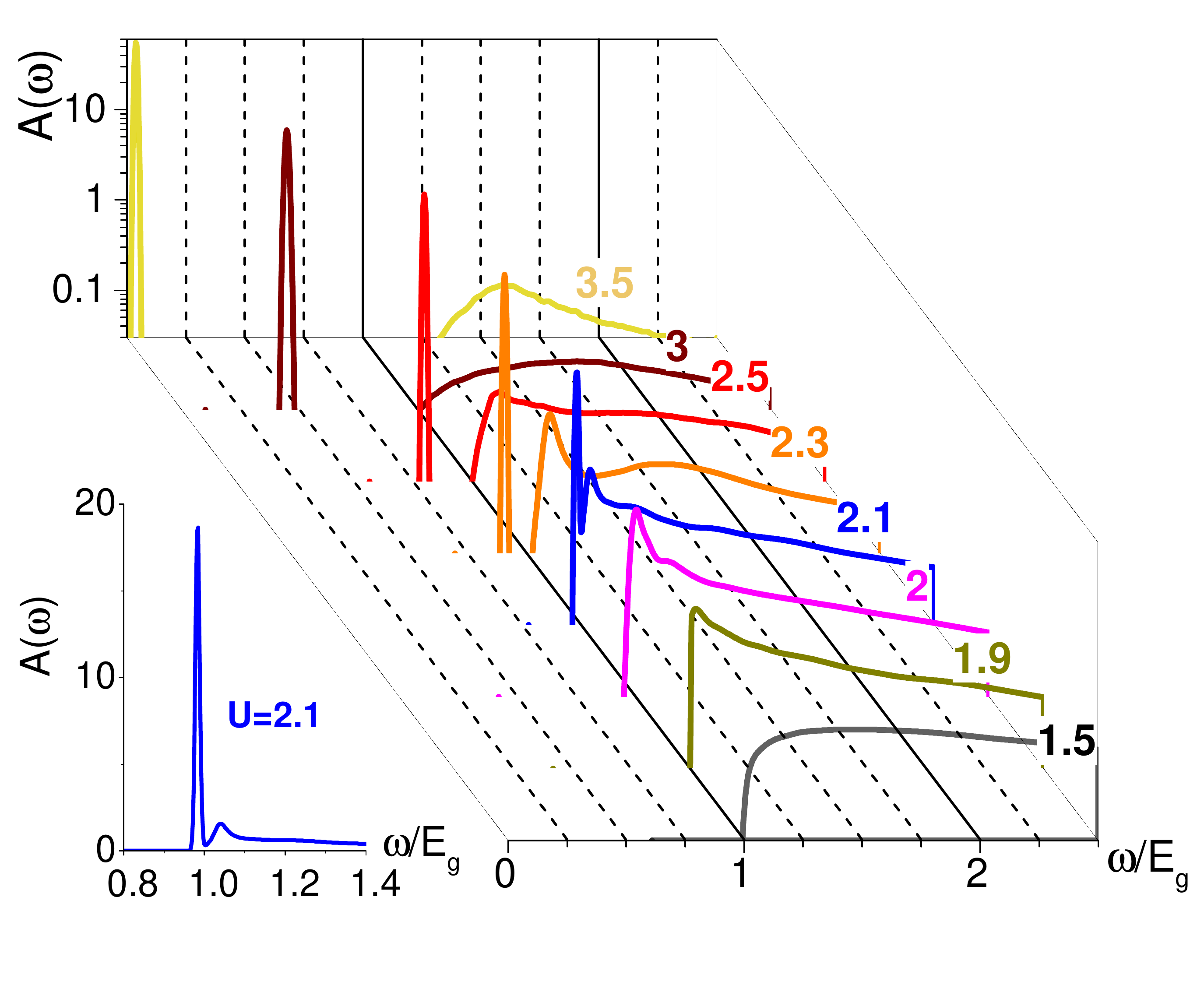}
\caption{(color online) 
Spectral function $A(\omega)$ for the pure excitonic model 
($\lambda_e=\lambda_h=0$) at different  $U$. 
Inset in the left shows $A(\omega)$ at $U=2.1$ in linear scale.}
\label{fig:fig_1}
\end{figure}

Our starting point is the optical response of the Coulomb-only 
interacting electron-hole system. 
From  Fig.~\ref{fig:fig_1} one clearly sees (i) the optical absorption threshold $\omega\simeq E_g$ 
for small $U$, (ii) how spectral weight accumulates at the bottom of the optical spectra as 
$U$ increases, (iii) that an excitonic peak splits off at the critical value $U_c\simeq 1.98$, 
which (iv)  separates more and more from the broad absorption band until it requires zero 
excitation energy  for very large Coulomb attractions when the exciton level approaches 
the valence-band top. 

\begin{figure}[bt]
\includegraphics[scale=0.45,clip=]{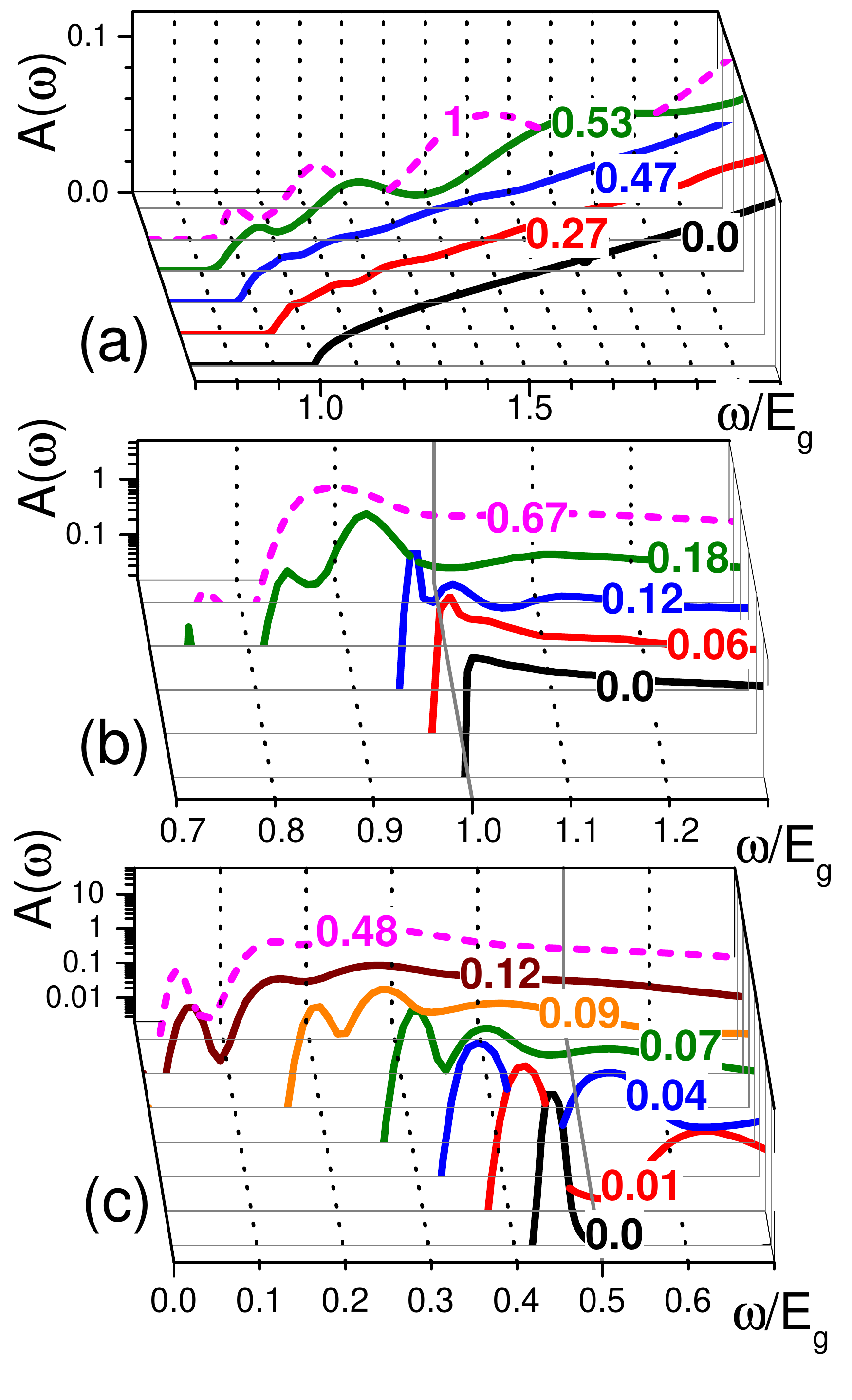}
\caption{(color online) 
Optical response depending on the fermion-phonon interaction strength (given at spectra) 
in the adiabatic regime $\Omega/E_g=0.1$, where 
$U/E_g=0$ (a), $U/E_g=1.9$ (b), and $U/E_g=3.0$ (c).
Results for  $\lambda_{e}=\lambda_{h}$ are given by solid lines; dashed magenta lines 
indicate the typical behavior when $\lambda_{e} > 0$ and $\lambda_{h}=0$.}
\label{fig:fig_2}
\end{figure}

\subsection{Coupling to the lattice: Adiabatic case}

Here we first neglect the Coulomb interaction between 
electrons and holes ($U=0$).  Results are inserted in Fig.~\ref{fig:fig_2}~(a). 
Working in the adiabatic regime, we observe (i) the lowering of the optical absorption 
threshold as the fermion-phonon coupling $g_e=g_h$ increases and (ii), at larger values 
of $\lambda$, an undulated absorption signal related to multi-phonon-involved processes 
when polaron formation sets in.  
Adding the coupling to the lattice degrees of freedom to the Coulomb attraction,  
in comparison to Fig.~\ref{fig:fig_1}, electron-hole bound states develop at substantially 
lower values of $U$ [note that $U<U_c$ in Fig.~\ref{fig:fig_2}~(b)]. 
Now the quasiparticle formed is largely dressed by phonons however, and therefore can be viewed 
as an exciton-polaron. 
The polaron exciton is characterized by a low spectral weight of the electronic part of the 
optical response because the electron-hole bound  state is entangled with a many-phonon (cloud) state.  
Then the small low energy peak at $\lambda=0.18$ in Fig.~\ref{fig:fig_2}~(b), 
separated from the rest of the spectra, is what 
is left projecting the exciton-polaron signal to a zero-phonon state. 
In accordance with that, the preexisting strong excitonic peak drawn  for $U=3$ and $\lambda=0$ [see Fig.~\ref{fig:fig_2}~(c)] is not only shifted to lower energies 
but lowered in intensity when the fermion-phonon coupling comes in to play. 
Intensity of low energy peaks in  Fig.~\ref{fig:fig_2}~(b-c) is lost to a series of  
phonon sidebands higher in energy, separated by the phonon frequency $\Omega=0.1E_g$. 
This  can be taken as a signature for exciton-polaron formation due to the constructive interplay 
of Coulomb and fermion-phonon interaction. Hence notable polaronic effects are observed  
even in the weak fermion-phonon coupling regime. 
We exemplarily included in Fig.~\ref{fig:fig_2} the case where the coupling to phonons takes place 
in the conduction band only (see dashed lines). 
Despite that the fermion-phonon interaction has to be approximately twice (four times) 
as large to form exciton-polarons for $U=0$ ($U\simeq U_c$), i.e., when a conduction 
band polaron catches  the hole in the valence band  developing a bound state, 
the spectra do not change much qualitatively.   
Let us emphasize that pattern of phonon sidebands appearing in Fig.~\ref{fig:fig_2} is  observed in the optical spectra in a whole range of different materials; see, e.g., Ref.~\onlinecite{QDST,s1,s2,s3,s4}.

\begin{figure}[t]
        \includegraphics[scale=0.45,clip=]{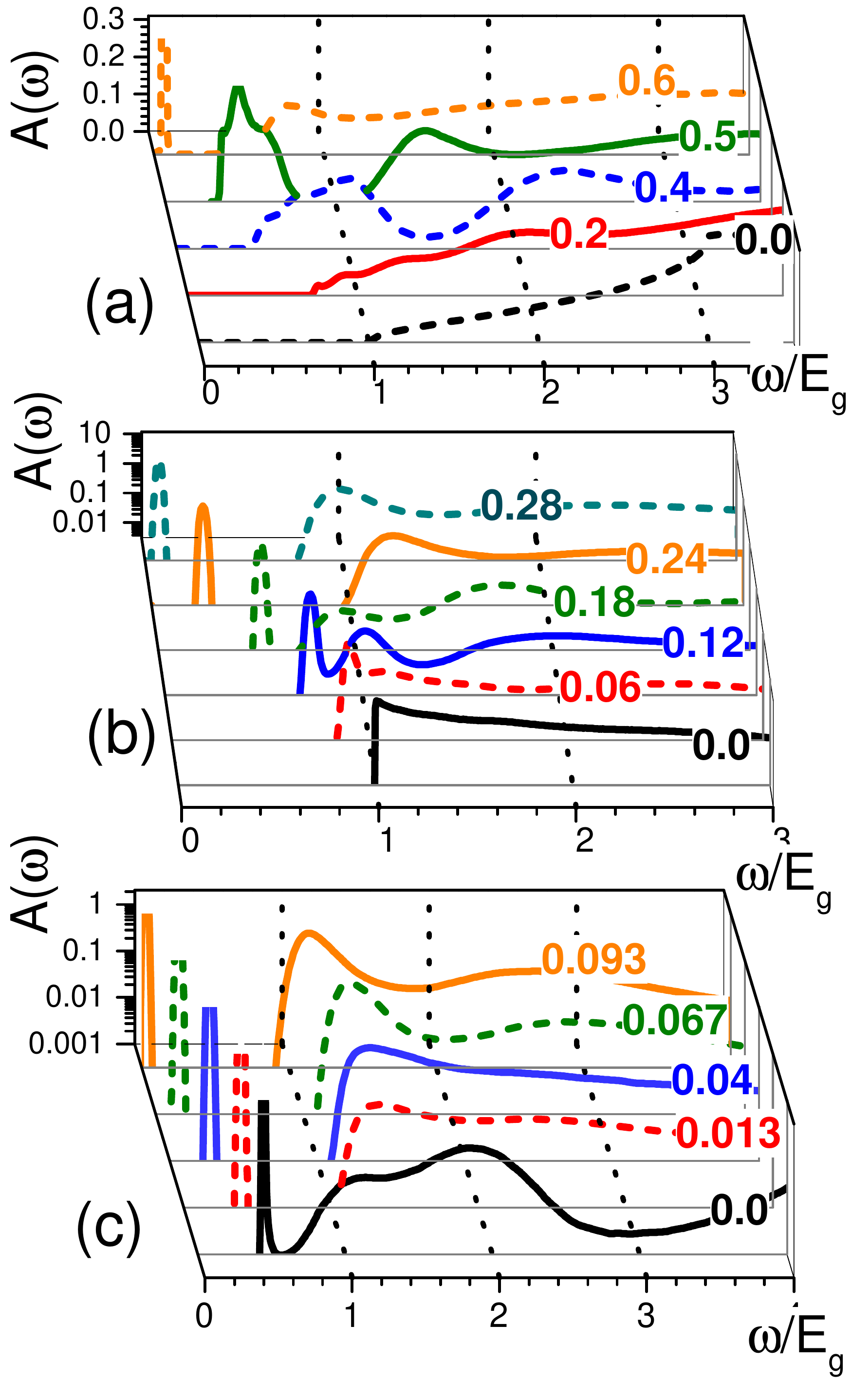}
\caption{(color online) 
Spectral function $A(\omega)$ in the nonadiabatic regime $\Omega/E_g=1$ at various $\lambda=\lambda_{e,h}$ (given at spectra)
for (a) $U=0$, (b) $U=1.9$, and (c) $U=3.0$.}
\label{fig:fig_3}
\end{figure}

\subsection{Coupling to the lattice: Non-adiabatic case.}

Let us now take a closer look  at the non-to-antiadibatic regime. 
Figure~\ref{fig:fig_3} presents the optical response for $\Omega/E_g=1$, i.e., the energy for bridging the band gap by photon absorption is the same as exciting one phonon.   Recall that the polaron crossover is much smoother in the antiadiabatic regime than in the adiabatic one, and a stronger fermion-phonon coupling $g_{e,h}$
is necessary because now the ratio between the polaron binding energy and the large frequency of the phonons matters~\cite{WF98,AFT10}. 
Large $\Omega$ is reflected in Fig.~\ref{fig:fig_3}~(a) for  $U=0$ in modulation of the spectral weight 
at the large  phonon scale ($\Omega=1$).
Due to resonance  $\Omega=E_g$ condition the low-energy peak, which can be attributed to an `bipolaronic' 
(electron-hole) quasiparticle bound by fermion-phonon interaction only~\cite{expo2008,Macrid2004},
becomes separated from the rest of the spectrum not at $\lambda=0.5$, as predicted in
 \cite{expo2008,Macrid2004}, but only when  $\lambda=0.6$. 
For $\lambda=0.5$, the lowest bump consists of three overlapping absorption signatures, separated by energies less than the phonon frequency. This is because calculating $A(\omega)$ we have to integrate over the momenta and, in the transition region, the bipolaron splits off from two (conduction and valence) bands having a renormalized (but still finite) bandwidth each, which leads to a central peak and a lower (upper) sattelites when the bipolaron band develops. 
By contrast, when the fermion-phonon interaction is present in only one of bands, we observed only a two-peak structure in the crossover regime (not shown).    
Figures~\ref{fig:fig_3}~(b) and (c) with $U/E_g=1.9$ and  $U/E_g=3.0,$ respectively,  illustrate the development of exciton-polarons in the  antiadibatic regime. Here the fermion-phonon coupling  $\lambda$ again triggers the formation of a  bound state for $U<U_c$ (see the curves for $\lambda= 0.18$, 0.24 and 0.28 in (b)), and a notable shift of the excitonic level towards the top of the valence band for  $U >U_c$  
[compare, in particular, Fig.~\ref{fig:fig_1} and \ref{fig:fig_3}~(c)].
A similar scenario is observed when only the electrons in the conduction band couple to phonons whereby, in this case, a larger (one-band) electron-phonon interaction is necessary. 
The nonadiabatic situation when the phonon frequency is comparable with the exciton binding energy, 
is for sure realized in many ionic crystals such as ZnO, MgO, BeO, TlCl, and TlBr. 
Here, in accordance with the corresponding experimental data~\cite{414a,414b,414c,460a,460b,461} 
the energy separation of the first phonon sideband from the main exciton peak 
is less than the optical phonon energy $\Omega$ (see Fig.~\ref{fig:fig_3}).

\section{Diagrammatics for the current-current correlation function}

The spectral function $A(\omega)$ is obtained by analytic continuation 
of the imaginary time current-current correlation function
$\Pi(\tau)=\langle \mathrm{vac} |  T_{\tau} \mathbf{j}(\tau)\mathbf{j}(0) 
| \mathrm{vac} \rangle$ 
to real frequencies,  solving the equation
$\Pi(\tau)=\int_{0}^{\infty} d\omega \exp(-\tau\omega)A(\omega)$
by the stochastic optimization consistent constraint method \cite{MPSS,Julich,CC,SOCC}.
The current operator, in real space, is defined as $\mathbf{j}= i [H,\mathbf{P}]$ with the  
polarization operator 
$\mathbf{P} = - e \mathbf{D} \sum_j e^{\dagger}_j h^{\dagger}_j + {\rm H.c.}$, where 
$\mathbf{D}=\int d \mathbf{r} \phi_e^*(\mathbf{r}) \mathbf{r}
 \phi_h(\mathbf{r})$  is the interband electron-hole dipole 
matrix element \cite{Mahan_Book}. For the model~\eqref{model}--\eqref{dispersion} under study, 
we obtain (e.g., in $x$-direction):
$j_x = j_x^{({f})} + j_x^{({p})}$, i.e. a sum of an purely
fermionic contribution,  
\begin{equation}
j_x^{({f})} = i e {D}_x \sum_{\bf k} f({\bf k}) X_{\bf k} + {\rm H.c.}\,,
\label{jf}
\end{equation}
and a polaronic contribution,
\begin{equation}
j_x^{({p})} = \frac{i e {D}_x}{\sqrt{N}} (g_e+g_h) \sum_{\bf kq}   Y_{{\bf k},{\bf q}} + {\rm H.c.}\,,
\label{jp}
\end{equation} 
where 
$X_{\bf k}=c_{\bf k}^{\dagger} h_{\bf -k}^{\dagger}$, $Y_{{\bf k},{\bf q}}=c_{\bf k}^{\dagger} h_{\bf -k-q}^{\dagger} (b_{\bf q}^{\dagger} + b_{\bf -q})$, and $f({\bf k}) = -(W_c+W_v)/6\sum_{\alpha} [1-\cos(k_\alpha)] - E_g + U$. 
For the limiting case of wide-gap semiconductor ($E_g \to \infty$) $f({\bf k})=f_c$ is a constant
whose absolute value $|f_c|$ is larger than all other characteristic energies $W_c$, $W_v$, 
and the maximum $\omega_\qq)$.
The detailed functional form of $f({\bf k})$ only marginally affects the results for the spectral function
and $f({\bf k})=f_c$  is set to unity (as well as electron charge, matrix element $|D_x|$, and 
Planck constant $\hbar$) for the calculations presented above.
Also, we will first neglect the polaronic contribution to the current and discuss its influence separately below.

\begin{figure}[t]
        \includegraphics[scale=0.28,clip=]{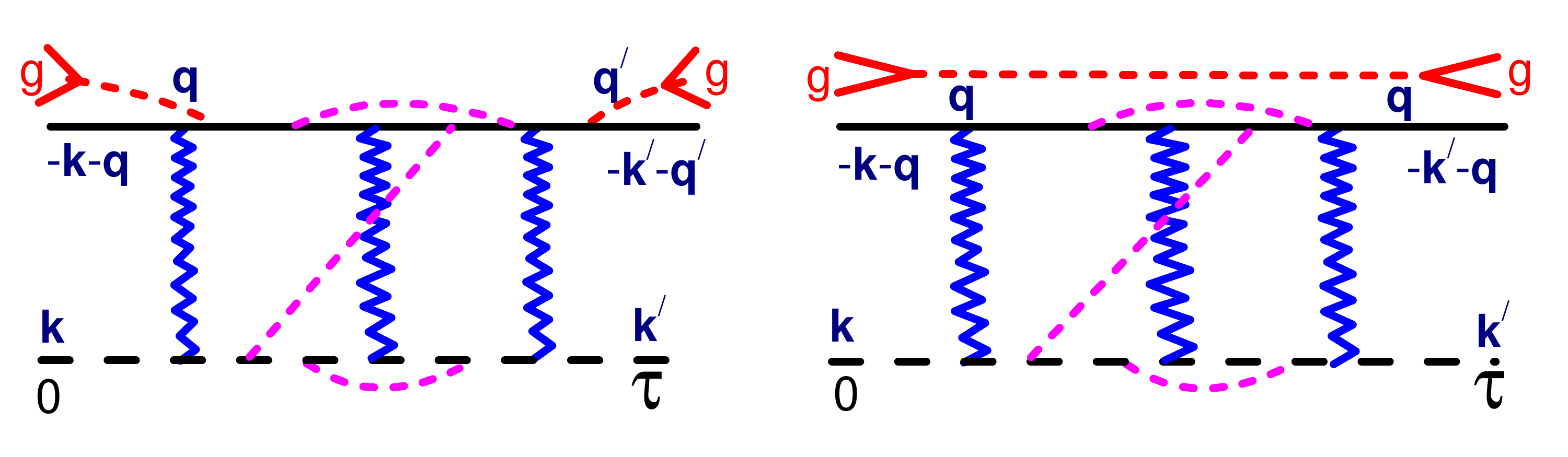}
\caption{(color online) 
Typical diagrams in the Feynman expansion of the current-current correlations function. Here solid (dashed) 
horizontal black lines represent electron (hole) propagators, dotted magenta lines denote phonon propagators, and 
wiggled blue lines symbolize the Coulomb attraction between electrons and holes. 
According to the absence (presence) of 
zero, one, or two phonon propagators attached  at imaginary times 0 or $\tau$ 
(red dotted line marked by 'g'), we obtain four different topologies of the left digram.
Presence (absence) of g-propagator means  that the correlation function is terminated by the operator 
$X_{\mathbf{k}}$ ($Y_{\mathbf{k},\mathbf{q}}$).}
\label{fig:fig_4}
\end{figure}

Adapting the DMC method to the calculation of the optical spectrum $A(\omega)$,
we rewrite the current-current correlation function in interaction representation and 
expand it with respect to both Coulomb attraction $(U)$ and fermion-phonon coupling  $(g_{e,h})$
strengths. Typical phonon-dressed ladder-type Feynman diagrams~\cite{FetterWalecka}, which contribute in such a series expansion, 
are depicted in   Fig.~\ref{fig:fig_4}. The weight attributed to a given diagram is the product of the interaction
vertices [$U(\pp,\kk,\kk')$,  $g_e(\qq)$, and $g_h(\qq)$] and Matsubara Green functions (for electrons, holes, and phonons) 
at imaginary times ($\tau$), where momentum conservation is imposed by the Hamiltonian~\eqref{model}.
We point out that the DMC updates of Coulomb vertices and phonon propagators are similar to those used for
the pure exciton~\cite{exciton2001,expo2008} and polaron~\cite{MPSS} problem, respectively. 
The main difference between the previous and present DMC implementations is due to 
the two distinct current operator contributions, \eqref{jf} and\eqref{jp}.  Accordingly, the DMC 
now has to switch between four topological different classes of Feynman diagrams, namely  
$XX$, $XY$,  $YX$, and  $YY$, depending on whether the beginning or end of the current-current correlation function 
is terminated by  $X$ or $Y$ operators, see Fig.~\ref{fig:fig_4}. The new DMC approach was validated, for the one-dimensional case, using expressions~\eqref{jf}-\eqref{jp}, by comparison with a truncated phonon-space exact diagonalization of~\eqref{model}\cite{LPBED,SSH,t-dep}. 
We work within DMC in the thermodynamic limit $N\to\infty$.

\begin{figure}[bt]
        \includegraphics[scale=0.30,clip=]{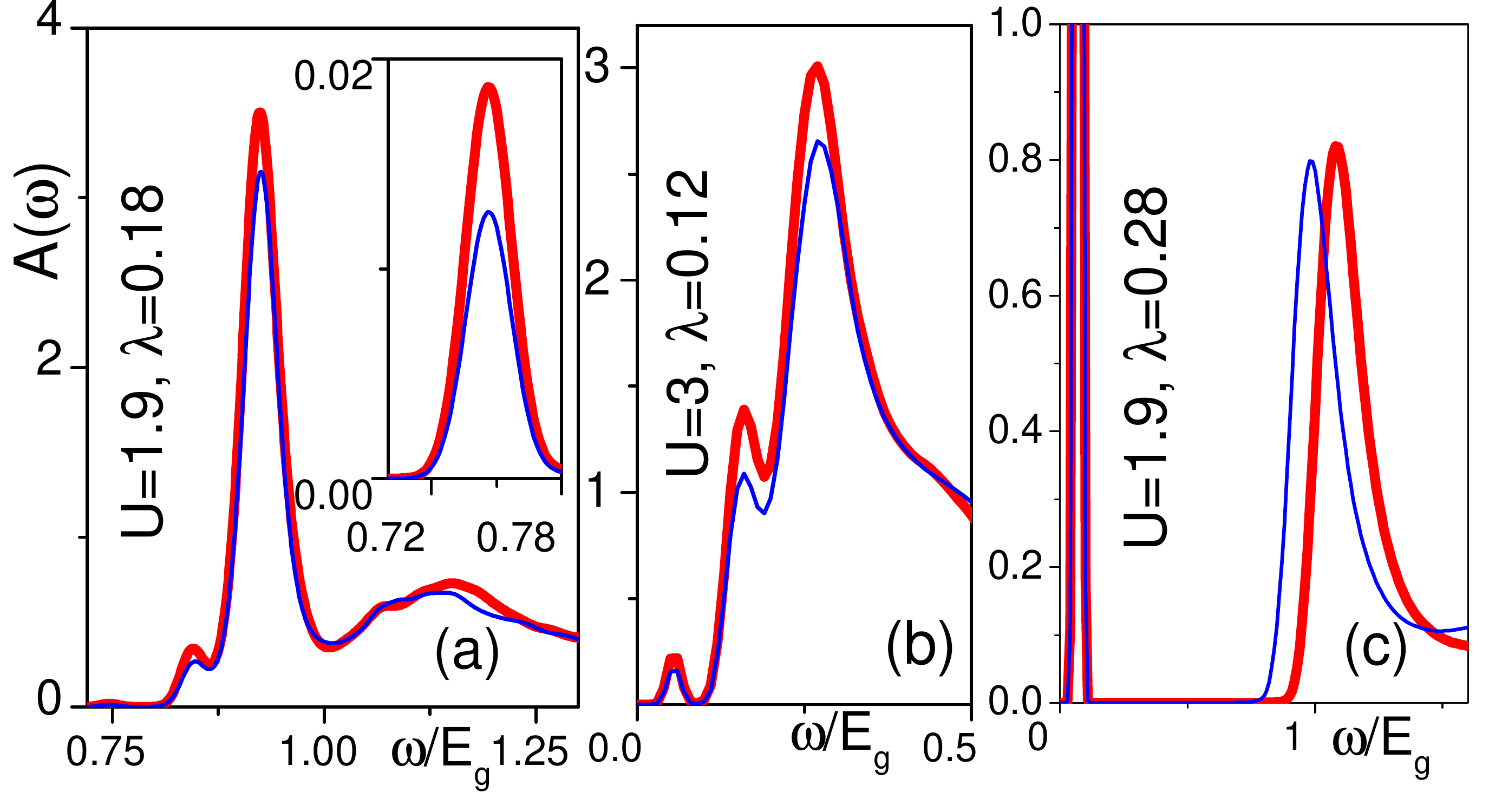}
     \caption{(color online) 
Contributions to the optical response, $A(\omega)$, in the adiabatic $\Omega/E_g=0.1$ [(a) and (b)] and antiadiabatic $\Omega/E_g=1.0$ (c) cases.
The  fermion-phonon couplings  are assumed to be the same in the valence and conduction bands, $\lambda_{e,h}=\lambda$.
Thick red (thin blue) lines give results when both current contributions  ${\bf j}^{(f)}$ and ${\bf j}^{(p)}$ (only ${\bf j}^{(f)}$) were included.} 
\label{fig:fig_5}
\end{figure}

Figure~\ref{fig:fig_5} illustrates the influence and relative importance of the electron 
${\bf j}^{({f})}$ and polaron ${\bf j}^{({p})}$ parts of the current to the spectral 
function $A(\omega)$. 
In the adiabatic regime, the polaronic current significantly contributes to the low-energy optical response of the system but the peak positions and the overall 
line shape are unaltered.  By contrast, in the antiadiabatic regime, the lowest peak, attributed to exciton-polarons, is less affected by ${\bf j}^{(p)}$ regarding the spectral weight; now ${\bf j}^{(p)}$ primarily influences the threshold and shape of the upper absorption band.

\section{Conclusions}

To sum up, utilizing a refined diagrammatic Monte Carlo method, we  discussed the optical absorption of photons by exciton-polarons within a paradigmatic two-band model, treating Coulomb and fermion-phonon coupling effects on an equal footing. The results are unbiased, and were derived for the infinite system in three dimensions.
Even though both interactions promote the formation of electron-hole bound states in the band gap, the resulting quasiparticles--and consequently their optical signatures--can be very different depending on the relative strength of the Coulomb attraction and the coupling to the lattice degrees of freedom where also retardation effects play an important role. While the predominantly Coulomb-bound exciton-polaron shows a strong and narrow resonance, the optical signal of a exciton heavily dressed by phonons has much low spectral weight, similar to bipolarons. When the phonons are slow (adiabatic regime), the low-energy contribution resulting from polaronic part of the current-current correlation function is substantial, whereas for fast phonons, in the non-to-antiadiabatic regime, the polaronic current mainly affects the threshold and lineshape of the upper optical absorption band. Our model calculations will support the analysis of optical absorption measurements in a wide class of correlated materials. 
   
\section{Acknowledgements}   
   
This work was supported by the ImPACT Program of the Council for Science, Technology and Innovation 
(Cabinet Office, Government of Japan), and by the Deutsche Forschungsgemeinschaft through SFB 652.


\begin{thebibliography}{99}

%
\bibitem{ExpOA} {\it Handbook of Spectroscopy}, ed. by G. Gaiglitz and 
T. Vo Dinh (VILLEY-VCH Verlag GmbH \& Co. KGaA, Weinheim, 2003).

\bibitem{Toyozawa_Book} Y. Toyozawa, \textit{Optical Processes in Solids}
           (Cambridge University Press, Cambridge 2003).

%
\bibitem{Egri85}  I.~Egri, Phys. Rep. \textbf{119}, 364 (1985).

%
\bibitem{Ueta_Book} M.~Ueta, H.~Kanzaki,
K.~Kobayashi, Y.~Toyozawa, and E.~Hanamura, \textit{Excitonic
Processes in Solids}, (Springer-Verlag, Berlin 1986).

%
\bibitem{Bardeen_Shockley_1950} J. Bardeen and W. Shockley, 
	           Phys. Rev. {\bf 80}, 72 (1950). 

%
\bibitem{Def2} M. V. Fischetti and S. E. Laux, 
                            J. Appl. Phys. {\bf 80}, 2234 (1996).

%
\bibitem{Def3} S.-H. Wei and A.  Zunger, 
                           Phys. Rev. B {\bf 60}, 5404 (1999).

%
\bibitem{Fomin98}  V. M. Fomin, V. N. Gladilin, J. T. Devreese, E. P. Pokatilov, S. N. Balaban, and S. N. and Klimin,
 \PRB \textbf{57}, 2415 (1998). 

%

\bibitem{VFB02} O. Verzelen, R. Ferreira, and G. Bastard, 
                     \PRL \textbf{88}, 146803 (2002).         

%
\bibitem{Wi83}  E.~G. Wilson, J. Phys. C: Solid State Phys. \textbf{16}, 1039 (1983).

\bibitem{HR68} B. I. Halperin and T. M. Rice, Rev. Mod. Phys. \textbf{40}, 755 (1968).     


\bibitem{PBF13}   V.-N. Phan, K. W. Becker, and H. Fehske,
                     \PRB \textbf{88}, 205123 (2013).     
                    

\bibitem{ZFB14}   B. Zenker, H. Fehske, and H. Beck,
                     \PRB \textbf{90},  195118 (2014).       
                     
   \bibitem{ZFBMB13}   B. Zenker, H. Fehske, H. Beck, C. Monney, and A. R. Bishop,
                     \PRB \textbf{88},  075138 (2013).       


\bibitem{KTKO13} T. Kaneko, T. Toriyama, T. Konishi, and Y. Ohta, \PRB
\textbf{87}, 035121 (2013).

\bibitem{KMCC02}
T. E. Kidd, T.  Miller, M. Y. Chou, and T. C Chiang, \PRL \textbf{88}, 226402 (2002).

\bibitem{WBM04}   P. Wachter, B. Bucher, and J. Malar,
                     \PRB \textbf{69},  094502 (2004).      
                     
 \bibitem{WB13} P. Wachter and B. Bucher, Physica B {\bf 408}, 51 (2013).    

\bibitem{QDST} A. S. Mishchenko and  N. Nagaosa,
                             Phys. Rev. Lett., {\bf 86}, 4624 (2001). 

\bibitem{RP} A. S. Mishchenko, N. Nagaosa, N. V. Prokof'ev, A. Sakamoto and B.V.~Svistunov,
                      Phys.\ Rev. B {\bf 66}, 020301(R) (2002).
 %

\bibitem{Sumi} A. Sumi, J. Phys. Soc. Japn. \textbf{43}, 1286 (1977). 
  
 \bibitem{PS98}
 N. V. Prokof'ev and B. V. Svistunov, 
                               \PRL \textbf{81}, 2514 (1998). 

\bibitem{MPSS}  A. S. Mishchenko, N. V. Prokof'ev, A. Sakamoto, 
                     and B. V. Svistunov,     \PRB \textbf{62}, 6317 (2000)
                                          
%
\bibitem{exciton2001}  E. A. Burovski, A. S. Mishchenko, 
                               N. V. Prokof'ev, and B. V. Svistunov, 
                               \PRL \textbf{87}, 186402 (2001). 
%
\bibitem{expo2008}  E. Burovski, H. Fehske, and A. S. Mishchenko, 
                               \PRL \textbf{101}, 116403 (2008). 
%
\bibitem{Mahan_Book} G. D. Mahan, \textit{Many Particle Physics}, 
                            (Plenum Press, New York 1990).
%
\bibitem{s1} R. M. Hochstrasser and P. N. Prasad, 
                        J. Chem. Phys. {\bf 56}, 2814 (1972).
%
\bibitem{s2} S. J. Xu, G. Q. Li, S.-J. Xiong, S. Y. Tong, C. M. Che, W. Liu, and M. F. Li,
                      J. Chem. Phys. {\bf 122}, 244712 (2005).
%
\bibitem{s3} S. J. Xu, G. Q. Li, S.-J. Xiong,  C. M. Che,
                       J. Appl. Phys. {\bf 99}, 073508 (2006)
%
\bibitem{s4} A. Brillante and M. R. Philpott,
                      J. Chem. Phys. {\bf 72}, 4019 (1980).     
%
 \bibitem{WF98} G. Wellein and  H. Fehske, 
           Phys. Rev. B {\bf 58}, 6208 (1998).
                   
%
 \bibitem{AFT10} A. Alvermann, H. Fehske, and S. A. Trugman, 
           Phys. Rev. B {\bf 81}, 165113 (2010).
           
           \bibitem{Macrid2004} A. Macridin, G. A. Sawatzky, and M. Jarrell, 
                Phys. Rev. B {\bf 69}, 245111 (2004).
%
\bibitem{414a} W. Y. Liang and A. D. Yoffe, 
                            \PRL \textbf{20}, 59 (1968). 
%
\bibitem{414b} R. C. Whited and W. C. Walker, 
                            \PRL \textbf{22}, 1428 (1969).
%
\bibitem{414c} W. C. Walker, D. M. Roessler, and E. Loh,
                         \PRL \textbf{20}, 847 (1968).
%
\bibitem{460a} R. Z. Bachrach and F. C. Brown, 
                          \PRL \textbf{21}, 685 (1968). 
%
\bibitem{460b} R. Z. Bachrach and F. C. Brown, 
                           Phys.\ Rev. B {\bf 1}, 818 (1970).
%
\bibitem{461} S. Kurita and K. Kobayashi, 
                         J. Phys. Soc. Jpn. {\bf 30}, 1645 (1971).
%
\bibitem{Julich} A. S. Mishchenko, in {\it Correlated Electrons: From Models to
                            Materials Modeling and Simulation}, edited by E. Pavarini, E.
                            Koch, F. Anders, and M. Jarrell (Verlag des Forschungszentrum,
                            J\"{u}lich, 2012), Vol. 2.
%
\bibitem{CC} N. Prokof'ev and B. Svistunov,
JETP Lett. {\bf 97}, 747 (2013).
%
\bibitem{SOCC} O. Goulko,  A. S. Mishchenko, L. Pollet, N. Prokof'ev, and B. Svistunov,
                            Phys. Rev. B {\bf 95}, 014102 (2017).
%

\bibitem{FetterWalecka} A. L.~Fetter and J. D.~Walecka,
                          \textit{Quantum Theory of Many-particle
                          Systems} (New York: McGraw-Hill, 1971).  
                          
%
\bibitem{LPBED} G. De Filippis, V. Cataudella, A. S. Mishchenko, 
           and N. Nagaosa, 
           Phys. Rev. B {\bf 85}, 094302 (2012).
           
           \bibitem{SSH} D. J. J. Marchand, G. De Filippis, V. Cataudella, 
                         M. Berciu, N. Nagaosa, N. V. Prokof.ev, 
                         A. S. Mishchenko, and P. C. E. Stamp, 
                         Phys. Rev. Lett. {\bf 105}, 266605 (2010).
           
           \bibitem{t-dep} G. De Filippis, V. Cataudella, E. A. Nowadnick, 
                     T. P. Devereaux, A. S. Mishchenko, and N. Nagaosa,
                     Phys. Rev. Lett. {\bf 109}, 176402 (2012).
 
 %
 \end{thebibliography}
\end{document}